\documentclass[final,5p,times,twocolumn]{elsarticle}

\usepackage{amssymb}
\usepackage{lipsum}
\usepackage{siunitx}
\usepackage{multirow}
\usepackage{float}
\usepackage{makecell}
\usepackage{amsmath}
\usepackage[colorlinks=true, allcolors=blue]{hyperref}

\journal{Results in Physics}

\begin{document}

\begin{frontmatter}



\title{Bayesian optimization of laser wakefield acceleration in the self-modulated regime (SM-LWFA) aiming to produce molybdenum-99 via photonuclear reactions}


\author[first,second,third]{Bruno Silveira Nunes\corref{cor1}}
\author[second,third]{Samara Prass dos Santos}
\author[fourth,third]{Roger Pizzato Nunes}
\author[fifth,sixth]{Cristian Bon\c{t}oiu}
\author[second,third]{Mirko Salomón Alva-Sánchez}
\author[first]{Ricardo Elgul Samad}
\author[first]{Nilson Dias Vieira Jr.}
\author[seventh,sixth]{Guoxing Xia}
\author[eighth]{Javier Resta-López}
\author[second,third]{Alexandre Bonatto\corref{cor1}}

\cortext[cor1]{Corresponding author.\\E-mail addresses: brunosnunes@usp.br (B. S. Nunes), abonatto@ufcspa.edu.br (A. Bonatto)}

\affiliation[first]{organization={Nuclear and Energy Research Institute, IPEN–CNEN},
            city={São Paulo},
            state={SP},
            country={Brazil}}
\affiliation[second]{organization={Graduate Program in Information Technology and Healthcare Management, Federal University of Health Sciences of Porto Alegre \\ (UFCSPA)},
            city={Porto Alegre},
            state={RS},
            country={Brazil}}
\affiliation[third]{organization={Beam Physics Group, Federal University of Health Sciences of Porto Alegre (UFCSPA)},
            city={Porto Alegre},
            state={RS},
            country={Brazil}}            
\affiliation[fourth]{organization={Federal University of Rio Grande do Sul (UFRGS)},
            city={Porto Alegre},
            state={RS},
            country={Brazil}}
 \affiliation[fifth]{organization={Department of Physics, University of Liverpool},
            city={Liverpool},
            country={United Kingdom}}
 \affiliation[sixth]{organization={The Cockcroft Institute, Sci-Tech Daresbury},
            city={Warrington},
            postcode={WA4 4AD},
            country={United Kingdom}}
 \affiliation[seventh]{organization={Department of Physics and Astronomy, University of Manchester},
            city={Manchester},
            country={United Kingdom}}
\affiliation[eighth]{organization={Instituto de Ciencia de los Materiales (ICMUV), Universidad de Valencia},
            city={Valencia},
            postcode={46071},
            country={Spain}}            

\begin{abstract}
While laser wakefield acceleration (LWFA) in the bubble regime demands ultra-short, high-peak-power laser pulses, operation in the self-modulated regime (SM-LWFA) works with more relaxed pulse conditions, albeit at the cost of lower beam quality. Modern laser systems can deliver pulses with durations of a few tens of femtoseconds and peak powers on the order of a few terawatts, at kHz repetition rates. These systems are well-suited for developing SM-LWFA applications where high average energy and charge are prioritized over beam quality. Such beams could be used to generate high-energy bremsstrahlung photons, capable of triggering photonuclear reactions to produce radioisotopes like molybdenum-99. This isotope decays into technetium-99m, the most widely used medical radioisotope, with over 30 million applications worldwide per year. This work explores the use of Bayesian optimization to maximize the energy and charge of electron beams accelerated via SM-LWFA. Particle-in-cell (PIC) simulations model a 5 TW, 15 fs-long Gaussian laser pulse, propagating through tailored hydrogen gas-density profiles. In these simulations, over multiple iterations, the algorithm optimizes a set of input parameters characterizing the gas-density profile and the laser focal position. Three distinct profiles, with total lengths ranging from 200 to 400 micrometers and combining ramps and plateaus, were investigated. Optimal configurations were found to produce electron beams with median energies ranging from 14 to 17 \si{MeV} and charges of 600 to 1300 \si{\pico\coulomb}, considering electrons with energies above 8 \si{MeV}. Preliminary estimates of the molybdenum-99 yields for the optimal beams were obtained by employing their phase spaces, retrieved from PIC simulations, as radiation source inputs in Monte Carlo simulations irradiating a combined tantalum and molybdenum target.
\end{abstract}




\end{frontmatter}




\section{\label{sec:intro}Introduction}

Plasma-based accelerators \cite{Esarey2009} can provide GV/m accelerating fields, driven by either laser pulses \cite{Tajima1979} or charged particle beams \cite{Chen1985} in plasmas. In both cases, as the driver propagates, it perturbs the plasma electron density in its wake, leading to the formation of high-amplitude wakefields that can be harnessed for charged-particle acceleration. Beam-driven wakefield accelerators (PWFA) have been employed as boosters to greatly increase the energy of a witness beam at the expense of the driver's energy loss \cite{Blumenfeld2007,DArcy2019}. Another distinctive application of PWFA has been implemented in the AWAKE experiment \cite{Adli2018, Gschwendtner2022}, in which a 400 \si{GeV} proton bunch drives a wakefield to accelerate electrons through a 10 m-long plasma. Nevertheless, while PWFA applications require a pre-accelerated beam driver, often generated by sizable accelerators, laser-driven wakefield accelerators (LWFA) have shown the capability of producing 8 \si{GeV} electron beams, self-injected from the plasma over a propagation distance of 20 cm \cite{Gonsalves2019}. Hence, laser-driven schemes might take greater advantage of the compactness anticipated in plasma-based accelerator applications.

Among the distinct regimes in which a LWFA can operate, the blowout or bubble regime \cite{Esarey2009} is known for being capable of producing high-energy electron beams, with low energy spread and emittance \cite{Gonsalves2019}. However, this regime typically requires laser pulses with high peak power, and lengths considerably shorter than the plasma wavelength. Such laser systems are costly, and often limited to operate at low repetition rates, in the sub-\si{\Hz} to \si{\Hz} domain. These properties may limit/restrict the use of the nonlinear regime for LWFA applications. On the other hand, the self-modulated regime (SM-LWFA), extensively explored in the 1990s \cite{Andreev:1992,Krall1993,Fisher1996} and later (with the advancement of laser technology) supplanted by the adoption of ultrashort pulses and the blowout regime, is now feasible with cost-effective laser systems at high repetition rates ($\sim$kHz). However, this comes at the expense of producing lower quality beams, with larger energy spreads and higher emittances if compared to the blowout regime \cite{Esarey2009,Maldonado2021}.

In the SM-LWFA, the laser self-focusing effect produces a near diffraction limited spot, enhancing nonlinearities that cause the self-modulation phenomenon, leading to laser-plasma resonant interaction. When the self-focusing phenomenon balances the diffraction for several plasma periods, acceleration of multi-MeV bunched electron beams can be obtained, albeit with broader energy spectra and greater divergences compared to the blowout regime. Despite the lower beam quality, the SM-LWFA can be achieved using longer laser pulses (compared to the plasma wavelength) with relatively low (sub-\si{\tera \watt} to few-\si{\tera \watt}) peak powers. Since pulses with these properties can now be obtained at high ($\sim$\si{\kHz}) repetition rates \cite{Rovige2020,Rovige2021}, the SM-LWFA might be suitable for applications that are not sensitive to the beam quality, such as the production of medical radioisotopes via photonuclear reactions. Among the possibilities, the production of molybdenum-99 ($^{99}$Mo, with a half-life of 66 \si{\hour}) stands out, as it subsequently decays into technetium-99m ($^\text{99m}$Tc, with a half-life of 6 \si{\hour}), the most widely used radioisotope in nuclear medicine, with over 30 million applications worldwide per year. Hence, obtaining $^{99}$Mo via photonuclear reactions could enable an alternative, highly-enriched-uranium-free production route for such radioisotope, thereby mitigating environmental and security concerns \cite{Nunes2022}.

$^{99}$Mo can be produced through the $^{100}$Mo($\gamma$,$n$)$^{99}$Mo reaction \cite{Chemerisov2012,Trknyi2018,Vieira2021}, whose cross-section is higher for photons with energies between 8 \si{\MeV} and 20 \si{\MeV}, peaking at approximately 14.5 \si{\MeV}. Bremsstrahlung photons with energies within this range could be generated by impinging a SM-LWFA-accelerated electron beam in a tantalum (Ta) converter, positioned prior to a Mo target. For this application, low-quality electron beams -- with high divergence and energy spread -- may suffice, provided they can generate a high flux of such bremsstrahlung photons. Due to the complex underlying physics of the SM-LWFA regime, the laser pulse dynamics as well as the dominant acceleration mechanisms strongly depend on the laser and plasma-density initial profiles. Hence, it is a non-trivial task to determine the optimal conditions that allow for producing electron beams with the highest energies and charge for a given scenario.

A useful approach for examining how initial laser and plasma conditions affect the acceleration of electron beams is to combine particle-in-cell (PIC) simulations \cite{Birdsall1991} with artificial intelligence algorithms. In particular, Bayesian optimization \cite{Brochu2010} can efficiently explore the extensive parameter space of laser and plasma settings, identifying conditions that may lead to improved outcomes towards in terms of the charge and energy spectrum of the electron beams. Bayesian optimization is known for its capability to handle functions with numerous inputs and unknown functional forms. It operates by incrementally building a dataset of function values, each new entry informing the next step in the search process. By using this iterative method, the algorithm can more effectively suggest parameter settings, even in scenarios with complex or less understood underlying relationships. This approach provides a balanced and steady progression towards finding more optimal configurations of laser and plasma parameters.

Bayesian optimization has recently been applied in two distinct LWFA experiments, in which the algorithm was used to control the parameters of a laser-plasma accelerator in  laboratory \cite{Shalloo2020,Jalas2021}. In addition, a study demonstrated the use of multitask Bayesian optimization \cite{Swersky2013} in PIC simulations, in which high- and low-fidelity simulations were used together to accelerate the optimization of a laser-plasma accelerator \cite{Pousa2022}.

In this study, Bayesian optimization is employed to determine the optimal values of multiple input parameters, characterizing the gas-density profile (target) and laser focal position in particle-in-cell (PIC) simulations modeling a SM-LWFA. These optimal values correspond to those maximizing an objective function based on the charge and energy spectrum of the accelerated electron beams, obtained as outputs from PIC simulations. This study is structured as follows. Section \ref{sec:model} describes the physical model, detailing the adopted PIC simulation parameters and providing an overview of the Bayesian Optimization algorithm employed. Sections \ref{sec:results1}, \ref{sec:results2}, and \ref{sec:results3} introduce three distinct hydrogen gas-jet density profiles, referred to as cases 1, 2, and 3, as well as the corresponding Bayesian optimization results. The optimized density profiles for each case are then compared in Section \ref{sec:comparison}. Finally, in section \ref{sec:conclusions}, the conclusions and future directions are discussed.

\section{\label{sec:model}The Model}

The physical system simulated in this work consists of a Gaussian laser pulse propagating through a transverse-homogeneous and longitudinally-varying density distribution of non-ionized hydrogen, simulating targets that could be obtained by using supersonic gas jets \cite{Schmid2012}, either in continuous flow \cite{He2013} or powered by rapid valves, enabling operation at frequencies on the order of kilohertz (\si{\kHz}). While the laser initial peak power, pulse duration, and transverse size at focus (waist) are kept constant, its focal position along the target length is one of the parameters to be optimized. Moreover, to ensure the system remains in the SM-LWFA regime, the lower bound of the plasma-density domain to be investigated is defined in such way that its plasma wavelength is shorter than the laser pulse length (FWHM) associated with the chosen pulse duration.

Regarding the target geometry, three gas-density profiles are proposed and will be clearly defined and illustrated in the forthcoming figures: an asymmetric trapezoid (Case 1), a profile featuring two plateaus connected by a down-ramp (Case 2), and a profile also comprising two plateaus, connected by either an up-ramp or a down-ramp (case 3). The utilization of asymmetric de Laval nozzles \cite{Couperus2016, Chiomento2021}, featuring designs that are adjustable \cite{Lei2023}, modular \cite{Lei2024}, or staged \cite{Tomkus2024}, enables the creation of tailored asymmetric gas-density profiles with distinct combinations of up- and down-ramps and plateaus, such as those investigated in this work. The height (i.e., the plasma density) and length of a plateau affect the wakefield amplitude and the LWFA dephasing length~ \cite{Esarey2009}. Up-ramps, on the other hand, serve a dual purpose. They not only can increase the energy gain by extending the dephasing length but also can help control the collimation and emittance of LWFA-accelerated electron bunches~ \cite{Aniculaesei2019,Yu2015}. Conversely, down-ramps can cause the injection of a substantial quantity of electrons into the acceleration bubble, effectively increasing the bunch charge \cite{Zhang2015,Swanson2017}. Additionally, a plasma-density decrease diminishes the phase velocity of the plasma wave \cite{Fubiani2006}, affecting the dephasing length as well \cite{Benedetti2015}. Given the complex dynamics of the laser pulse in the SM-LWFA, predicting the optimal lengths and heights of such ramps and plateaus is not straightforward. Hence, in this work the parameter space is systematically explored using Bayesian optimization, to identify configurations that maximize the energy and charge of the LWFA-accelerated electron bunches. This approach allows for the most favorable conditions to be determined in a data-driven manner, significantly reducing the computational efforts required for optimizing the SM-LWFA.

\subsection{\label{sec:pic}PIC simulations}

PIC simulations were carried out using the Fourier-Bessel PIC (FBPIC) code \cite{Lehe2016}, which adopts a spectral solver that uses a set of 2D (RZ) grids, each of them representing an azimuthal mode $m$. Among the notable features of FBPIC are the mitigation of spurious numerical dispersion by its spectral solver algorithm, including the zero-order numerical Cherenkov effect~ \cite{Godfrey1974}, and the implementation of the openPMD metadata standard ~ \cite{Huebl2018}. The simulation domain, spatial resolution, number of azimuthal modes, and number of particles per cell adopted in the PIC simulations presented in this work are listed in Table \ref{tab:PIC}. To ensure reproducibility and minimize the effects of random fluctuations on the Bayesian optimization algorithm, a fixed random seed, arbitrarily selected, was adopted for all simulations, except where specified otherwise.

In all simulations, a 5 \si{\tera\watt} peak power laser pulse, linearly polarized in the $x$ axis, propagates (along the $z$ coordinate) through a density profile of non-ionized hydrogen. The laser pulse is defined as a Gaussian beam, focused to a waist of $w_0 = 7$ \si{\um} at a given focal position. As the laser propagates through the neutral-gas target, the local density of plasma electrons is calculated using the ADK ionization model~ \cite{Ammosov1986}. Besides being consistent with previous studies \cite{Salehi2017, Maldonado2021, Goers2015, Woodbury2018}, these parameters are compatible with the design of a SM-LWFA under development at the Center for Lasers and Applications of the Nuclear and Energy Research Institute (IPEN-CNEN/SP, Brazil) \cite{Bonatto:2021}. Table \ref{tab:laser} provides the set of laser parameters adopted in all PIC simulations performed. The laser focal position and the target profile, which are the parameters to be optimized, will be introduced in the following sections.

\begin{table}[!t]
\caption{\label{tab:PIC} PIC parameters.}
\centering
\begin{tabular}{@{}cccl}
\hline
\multirow{2}{*}{\makecell[c]{Parameter}} & \multirow{2}{*}{\makecell[c]{Value}} & \multirow{2}{*}{\makecell[c]{Unit}} &\multirow{2}{*}{\makecell[c]{Description}}\\
&&\\
\hline
  \multirow{2}{*}{$z_{min}$} &  \multirow{2}{*}{-100} & \multirow{2}{*}{\si{\um}} &
 \multirow{2}{*}{\makecell[l]{initial boundary of the\\longitudinal simulation domain}} \\
  &  & & \\[1.0ex]
 \multirow{2}{*}{$z_{max}$} & \multirow{2}{*}{0} & \multirow{2}{*}{\si{\um}} &  \multirow{2}{*}{\makecell[l]{final boundary of the\\longitudinal simulation domain}}\\
 &&&\\[1.0ex]
  \multirow{2}{*}{$r_{1}$} &  \multirow{2}{*}{20} &   \multirow{2}{*}{\si{\um}} &  \multirow{2}{*}{\makecell[l]{simulation domain radius\\(Case 1)}}\\
  &&&\\[1.0ex]
 \multirow{2}{*}{$r_{2} \,,\, r_{3}$} & \multirow{2}{*}{50} &  \multirow{2}{*}{\si{\um}} & \multirow{2}{*}{\makecell[l]{simulation domain radius\\(Cases 2 and 3)}}\\
 &&&\\[1.0ex]
 $\Delta_z$ & 27 & \si{\nm} & spatial resolution in $z$  \\[1.0ex]
 $\Delta_r$ & 33 & \si{\nm} & spatial resolution in $r$ \\[1.0ex]
  $N_m$ & 3 & --- & number of azimuthal modes \\[1.0ex]
 $N_{pz}$ & 2 & --- & particles per cell along $z$  \\[1.0ex]
 $N_{pr}$ & 2 & --- & particles per cell along $r$  \\[1.0ex]
 $N_{p\theta}$ & 12 & --- & particles per cell along $\theta$  \\        
 \hline
%
\end{tabular}

\end{table}

\begin{table}[!t]
\caption{\label{tab:laser} Laser parameters.}
\centering
\begin{tabular}{@{}cccl}\hline

\multirow{2}{*}{\makecell[c]{Parameter}} & \multirow{2}{*}{\makecell[c]{Value}} & \multirow{2}{*}{\makecell[c]{Unit}} &\multirow{2}{*}{\makecell[c]{Description}}\\
&&\\
\hline

$P_L$ &  5 & \si{\tera\watt} & initial peak power\\[1.0ex]
$\lambda_0$ & 800 & \si{\nm} & wavelength  \\[1.0ex]
$z_0$ & -50 & \si{\um} & pulse centroid\\[1.0ex]
 $c\tau$ & 15 & \si{\um} & pulse length (FWHM)\\[1.0ex]
$w_0$ & 7 & \si{\um} & waist\\[1.0ex]
$a_0$ &  1.74 & --- & strength parameter  \\
\hline
\end{tabular}
\end{table}

\subsection{\label{sec:bo}Bayesian Optimization}

In this study, the Bayesian optimization algorithm was implemented using the Botorch library \cite{Balandat2019}, aiming to maximize an objective function $F_\text{obj}$, to be defined in terms of the energy and charge of the LWFA-accelerated beam. Utilizing data from an initial set of simulations, performed with input parameters randomly chosen within their variation ranges, the algorithm constructs a surrogate model via Gaussian process regression \cite{Rasmussen2008}, in order to predict the behavior of the objective function. Given $x_1, ..., x_k$ as the values of the $k$ input parameters to be optimized, this model estimates the mean $\mu(x_1, ..., x_k)$ and standard deviation $\sigma(x_1, ..., x_k)$ of $F_\text{obj}$ at unexplored points, representing its expected value and associated uncertainty, respectively. These results are then utilized to evaluate an acquisition function, which in turn is used to infer a new set of input parameters for the subsequent PIC simulation. In this work, the Upper Confidence Bound (UCB) \cite{Srinivas2012} acquisition function has been adopted,
\begin{align}
\text{UCB} &= \mu(x_1, ..., x_k) + \sqrt{\beta_i} \, \sigma(x_1, ..., x_k) \, , \\
\beta_i &= 2\log{\frac{N(i\pi)^2}{6\delta}} \, ,
\end{align}
where $\beta_i$ is the trade-off parameter between mean and covariance~ \cite{Srinivas2012}, which adjusts the balance between exploring new areas and exploiting known regions. Although $\beta_i$ can be set as a constant, the functional form adopted here depends on the iteration number $i$, the total number of data points $N$, and a user-specified parameter $\delta$, which affects the level of confidence in exploration, selected within the range $0 < \delta \leq 1$ (in this work, $\delta = 1$ was adopted). Through iterative refinement, the Bayesian optimization algorithm efficiently explores the parameter space, progressively improving the surrogate model to identify the optimal parameters that maximize $F_\text{obj}$.

For each PIC simulation, the input parameters to be optimized are the ramps and plateaus specifications that define the adopted gas-density profile, alongside the laser focal position. Tables \ref{tab:case1}, \ref{tab:case2}, and \ref{tab:case3} list the input parameters associated with the three distinct profiles, cases 1, 2, and 3, respectively, to be examined in the following sections. Regarding the output parameters, they are extracted from the energy spectrum of the accelerated beam obtained from PIC simulation results. Each spectrum is built as a 200-bin histogram, being $E_i$ and $Q_i$ the kinetic energy and charge, respectively, associated with the $i$-th bin. Aiming to maximize the fraction of high-energy electrons, the following objective function is adopted,
\begin{equation}
    F_\text{obj} = \sum_{i}^{N=200} E_i \, Q_i \quad , \quad E_i \geq 8\text{ \si{\mega \electronvolt}} \,.
    \label{eq:Fobj}
\end{equation}
 In Eq.~(\ref{eq:Fobj}), the 8 \si{\MeV} threshold was adopted because electrons with lower energies will not produce photons capable of triggering the $^{\text{100}}$Mo($\gamma$,n)$^{99}$Mo reaction route \cite{Trknyi2018}. Moreover, this choice prevents the algorithm from optimizing the system by greatly increasing the charge of low-energy bins. Although in Eq. (\ref{eq:Fobj}) $E_i$ is computed in \si{\MeV} and $Q_i$ in \si{\nano\coulomb}, the objective function will be referred to as having an arbitrary unit (arb. unit, or a.u.). In addition, the energy and charge will be labeled as $E_{sel}$ and $Q_{sel}$, respectively, in quantities and figures obtained considering the 8~\text{\si{\MeV}} threshold. This notation ensures clarity in identifying analyses that are specific to the selected energy range.

\section{\label{sec:results1}Optimization of a trapezoidal gas-density profile\\(Case 1)}

\begin{figure}
    \centering
    \includegraphics{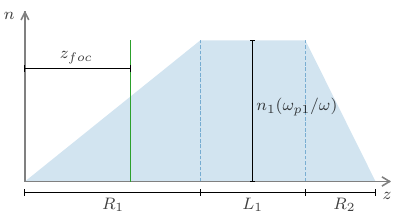}
    \caption{Trapezoidal gas-density profile (Case 1).}
    \label{fig:case1}
\end{figure}

The first investigated profile has a longitudinal trapezoidal shape, with up- and down-ramps free to have distinct, independently-optimized lengths, generalizing the symmetric profiles adopted in previous studies \cite{Maldonado2021,Aniculaesei2019,Sundstrm2022}  to model the gas flow of submillimetric supersonic nozzles \cite{Schmid2012,Couperus2016}. In this profile, illustrated in Fig.~\ref{fig:case1}, a linear up-ramp increases the density of hydrogen atoms from zero to $n_1$, over a length $R_1$. The density remains constant at $n_1$ along a plateau of length $L_1$, linearly decreasing to zero over a down-ramp of length $R_2$. Fig.~\ref{fig:case1} also illustrates the laser focal position, $z_{foc}$, and shows $n_1$ as a function of the plasma-to-laser frequency ratio, $n_1(\omega_{p1}/\omega)$. Since each hydrogen atom has a single electron, upon complete ionization, $n_1$ also represents the plasma electron density. Hence, a plasma frequency $\omega_{p1} = \sqrt{n_1 e^2/(\epsilon_0 \, m_e)}$, where $e$ is the elementary charge, $\epsilon_0$ is the vacuum permittivity, and $m_e$ is the electron rest mass, can be associated with $n_1$. Because this ratio characterizes the plasma's response to the laser pulse propagation (underdense for $\omega_{p1}/\omega < 1$, critical density at $\omega_{p1}/\omega = 1$, and overdense for $\omega_{p1}/\omega > 1$), it is adopted here as an input parameter rather than directly using the electron density $n_1$. Furthermore, for a given $\omega_{p1}/\omega$, the density $n_1$ can be readily calculated using the expression $n_1 = (\omega_{p1}/\omega)^2 (2 \pi c/\lambda)^2 (\epsilon_0 m_e/e^2)$, where $c$ is the speed of light and $\lambda$ is the laser wavelength. Table~\ref{tab:case1} specifies the ranges for optimizing $z_{foc}$, $\omega_{p1}/\omega$, $R_1$, $L_1$, and $R_2$, chosen based on initial simulations and typical values used in LWFA experiments and simulations with gas nozzle targets. The range interval for $\omega_p/\omega$ presented in Table~\ref{tab:case1} corresponds to gas densities ranging from $n = 0.4 \times 10^{20},\si{\per\cm\cubed}$ to $n = 2.8 \times 10^{20},\si{\per\cm\cubed}$, which can be experimentally obtained with de Laval nozzles.

\begin{table}

\caption{\label{tab:case1}Case 1: input parameters.}
\centering
\begin{tabular}{@{}cccl}
\hline
\multirow{2}{*}{\makecell[c]{Parameter}} & \multirow{2}{*}{\makecell[c]{Range}} & \multirow{2}{*}{\makecell[c]{Unit}} &\multirow{2}{*}{\makecell[c]{Description}}\\
&&\\
\hline
     $z_{foc}$ & $\phantom{0}20\;,\,160$& $\si{\um}$ & laser focal position\\[1ex]
        \multirow{2}{*}{$\omega_{p1}/\omega$} &  \multirow{2}{*}{$0.15 \;,\, 0.40$}& \multirow{2}{*}{---}  & \multirow{2}{*}{\makecell[l]{plasma-to-laser \\ frequency ratio}}\\
        &&&\\[1ex]

     $R_1$ & $\phantom{0}40 \;,\, 120 $&$\si{\um}$ & up-ramp length\\[1ex]
        $L_1$ & $ 40 \;,\, 80 $&$ \si{\um}$ & plateau length \\[1ex]
          $R_2$ & $ \phantom{0}40 \;,\, 120 $&$\si{\um}$ & down-ramp length\\ 
          \hline
\end{tabular}
\end{table}

\begin{figure}[!t]
    \centering
    \includegraphics{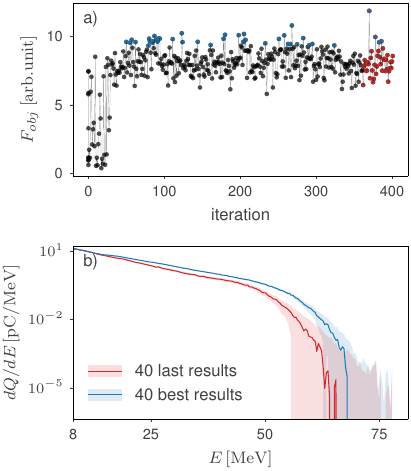}
    \caption{(a) $F_{obj}$ along the optimization of Case 1. Blue and red dots represent the 40 best and 40 last simulations, respectively. (b) Average energy spectra of the 40 best (blue line) and 40 last (red line) simulations, with shaded areas representing their 95\% confidence intervals.} 
    \label{fig:caso1_iter}
\end{figure}

\begin{figure}[!t]
    \centering
    \includegraphics[scale=0.97]{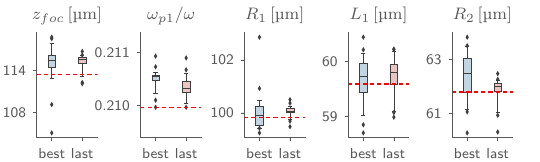}
    \caption{Input parameter distributions for the 40 best (blue boxplots) and 40 last (light-red boxplots) simulations of Case 1. The red dashed lines indicate the optimal values for each parameter.}  
    \label{fig:boxplots_caso1}
\end{figure}

Figure~\ref{fig:caso1_iter}(a) depicts the iterative evolution of the objective function over more than 400 PIC simulations, each conducted with input parameters defined by the Bayesian optimization algorithm. While $F_{obj}$ initially exhibits a wide range of values, after approximately 30 iterations it stabilizes around an average of $F_{obj} \simeq (8\pm 1)\, \si{a.u.}$. The absence of a further increasing trend, even after 400 simulations, suggests convergence of $F_{obj}$ to this value by the algorithm. Observed oscillations in $F_{obj}$ likely result from a balance between exploration and inherent simulation noise, which can include variability due to the stochastic nature of some FBPIC processes not necessarily governed by a fixed, user-defined random seed. Additionally, since particles are weighted by their energies in Eq.~\ref{eq:Fobj}, $F_{obj}$ is inherently sensitive to high-energy particles. Hence, minor changes in the fraction of such particles may significantly affect its value. In order to evaluate the significance of the $F_{obj}$ variation range after it has potentially converged, the energy spectra of two representative groups of 40 simulations  (approximately 10\% of the total number of simulations) are compared. The first group comprises the 40 simulations with the highest $F_{obj}$ values (40 best results), shown as blue dots in blue in Figure~\ref{fig:caso1_iter}(a), and the second group comprises the last 40 simulations (40 last results), shown as red dots in the same figure. 
Figure~\ref{fig:caso1_iter}(b) shows the median spectra for both groups alongside their 95\% t-student confidence intervals (colored areas). The 40 last features $F_{obj} = 6.6$ \si{{u.a.}} and $Q_{sel} \simeq 380 \,\si{pC}$, whereas the 40 best presents $F_{obj} = 8.8$ \si{{u.a.}} and $Q_{sel} \simeq 450 \,\si{pC}$, indicating a 30\% increase in the objective function and a 20\% rise in selected charge for the 40 best group. Despite a close match at lower energies, discrepancies widen at higher energies, with the of the 40 best simulations showing greater charge than the 40 last ones. While the maximum energy in the 40 best also surpasses the 40 last, the difference is not statistically significant, as evidenced by the considerable overlap of the confidence intervals beyond approximately 60 \si{MeV}. Despite the aforementioned differences, the distributions of the input parameters for both groups, shown in Figure~\ref{fig:boxplots_caso1}, suggest they are physically similar.

\begin{figure*}[!t]
    \centering
    \includegraphics{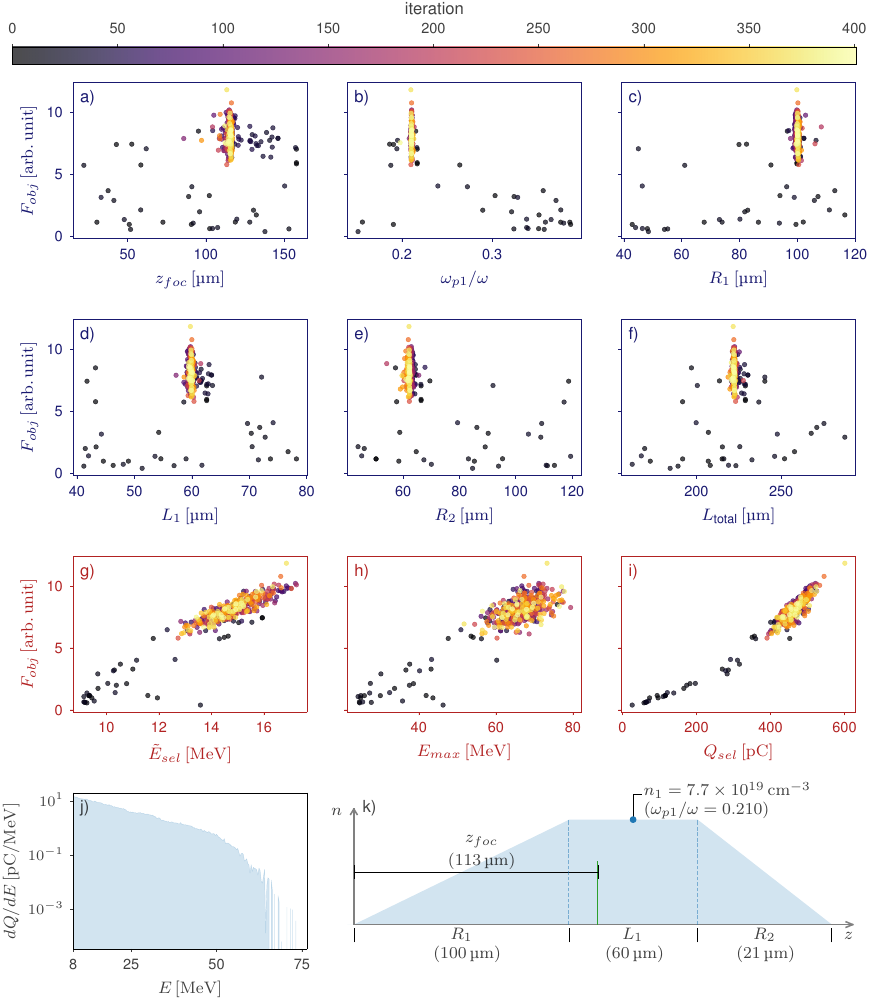}
    \caption{Bayesian optimization of Case 1, with a color scale indicating the algorithm iteration. Input parameters: $F_{obj}$ versus the (a) laser focal position, (b) plasma-to-laser frequency ratio at the plateau, and lengths of the (c) up-ramp, (d) plateau, and (e) down-ramp sections of the gas-density profile, along with the (f) total length of the profile. Output parameters: $F_{obj}$ versus the (g) median energy (above 8 \si{MeV}) $\tilde{E}_{sel}$, (h) maximum energy, and (i) selected charge $Q_{sel}$. Case 1 optimal: (j) energy spectrum and (k) gas-density profile, with the optimal values for the input parameters.}
    \label{results2}
\end{figure*}

Figures~\ref{results2}(a)-(f) present scatter plots that track the evolution of the objective function with respect to each input parameter listed in Table~\ref{tab:case1}, throughout the iterations of the Bayesian optimization algorithm. The total length of the gas-density profile, $L_{total}$, plotted in Fig.~\ref{results2}(f), although not listed in Table~\ref{tab:case1}, is the sum of the up-ramp, plateau, and down-ramp lengths. The color scale is associated with the iteration number, with lighter colors corresponding to later iterations. In all panels, clusters of light-colored points, particularly at higher values of  $F_{obj}$, suggest convergence of the algorithm for the respective input parameters. Similar plots are shown in figures \ref{results2}(g)-(i) for the output parameters, to characterize the accelerated electron beam in each simulation. Figure~\ref{results2}(g) shows the median energy above 8 \si{\MeV}, $\tilde{E}_{sel}$. In this panel, a cluster of light-colored points indicates that beams with median energies between 13 \si{\MeV} and 17 \si{\MeV}, and corresponding values of $F_{obj}$ between 6 \si{a.u.} and 10 \si{a.u.}, were produced following the convergence of the algorithm. Moreover, $\tilde{E}_{sel} = 16.9$ \si{\MeV} was obtained for the highest value of $F_{obj}$ achieved. Figure~\ref{results2}(h) presents the maximum energies ($E_{max}$) obtained in each simulation. Although this parameter is not statistically significant, as only a few electrons achieve such energies, it is interesting to observe that, during convergence, maximum energies between 60 \si{\MeV} and 80 \si{\MeV} were obtained, with $E_{max} = 73$ \si{\MeV} for the simulation with the highest value of $F_{obj}$ achieved. Figure~\ref{results2}(i) depicts the selected charge ($Q_{sel}$) of the beams, considering only particles exceeding 8 \si{MeV} in energy. Selected charges between 400 \si{\pico\coulomb} and 600 \si{\pico\coulomb} were achieved during convergence, with the highest value obtained in the optimal simulation, that maximized $F_{obj}$. Figures~\ref{results2}(g)-(i) depict a clear improvement due to optimization, with darker dots at lower parameter values transitioning to brighter, clustered dots at higher values.  Fig.~\ref{results2} is concluded with panels (j) and (k) presenting the optimal beam energy spectrum and gas-density profile, respectively.

\begin{figure*}[!t]
    \centering
    \includegraphics{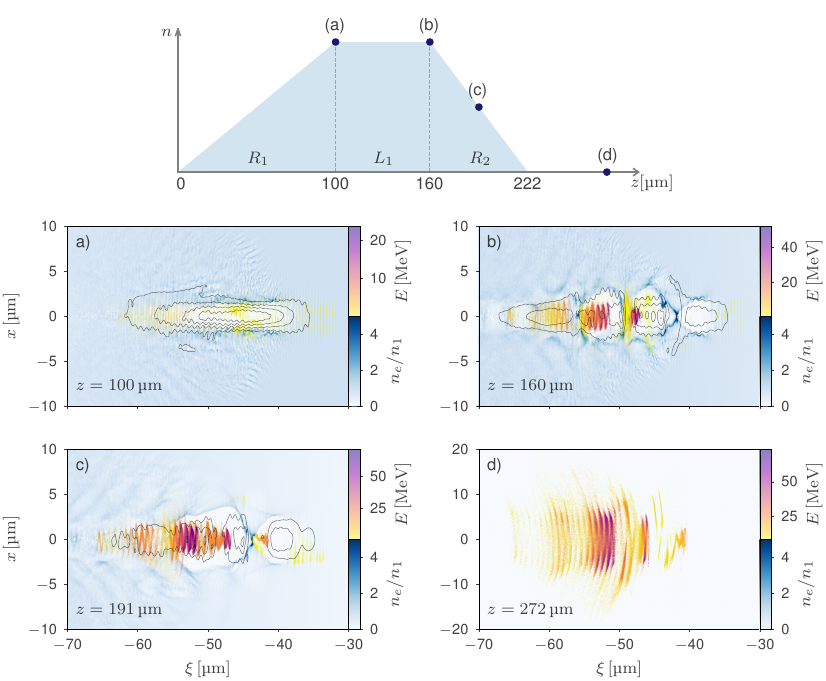}
    \caption{Spatial evolution of the optimal simulation, Case 1. The upper panel shows the propagation distances at which panels (a) to (d) were plotted. In these panels, the plasma electron density (normalized by the plateau density) is shown using a blue gradient scale, and the backtracked electrons, which will attain energies exceeding 8 \si{MeV} after ejection, are depicted as colored dots with a yellow-purple color scale representing their energies at the current propagation distance. Additionally, contour lines for $a_0$ ranging from 1 (outermost) to 4 (innermost), in unitary steps, depict the laser envelope at these distances.} 
    \label{nelec2}
\end{figure*}

The underlying physics of the optimal simulation of Case 1 can be assessed by backtracking the trajectories and energy evolution of the electrons that constitute the accelerated beam, specifically those exiting the gaseous target with energies exceeding 8 \si{MeV}. Figures~\ref{nelec2}(a)-(d) illustrate the plasma electron density normalized by the density of the first plateau, $n_e/n_1$, in a blue gradient scale. Backtracked electrons are depicted as colored dots, with a yellow-purple color scale representing their energies at each location. Additionally, contour lines for the laser strength parameter $a_0$, ranging from 1 (outermost) to 4 (innermost) in unitary steps, were added to show the laser envelope. Each panel is plotted at a distinct propagation distance $z$ along the gas-density profile, illustrated in the upper, unlabeled panel of Fig.~\ref{nelec2}. In the labeled panels, the moving coordinate $\xi \equiv z - v_w \, t$ is adopted, where $v_w$ is the velocity of the simulation moving window, chosen to approximately lock the laser pulse in the frame.

Figure~\ref{nelec2}(a) reveals a long and tightly focused laser pulse at the beginning of the density plateau ($z = 100 \, \si{\micro m}$), evident from the $a_0$ contour lines. Within the laser pulse, the intense fields cause a near-complete depletion of the normalized plasma electron density ($n_e/n_1$), visualized as a white region. Despite this depletion, a small fraction of the electrons that will contribute to the final accelerated beam, most of them still with energies below 10 \si{MeV} (yellow to orange dots), can be seen gaining kinetic energy within this region, likely through direct laser acceleration (DLA)~ \cite{Pukhov1999,Shaw2016,Wang2019}. Along the plateau, the laser pulse drives an intense wakefield (not shown in Fig.~\ref{nelec2}), with a wavelength shorter than the laser envelope. The interaction between the long pulse and the disturbed plasma leads to laser self-modulation \cite{Esarey2009}, a process where the laser intensity becomes longitudinally modulated within the pulse envelope. The $a_0$ contour lines in Fig.~\ref{nelec2}(b), plotted at the end of the plateau ($z = 160 \, \si{\mu m}$), depict laser pulse fragments with lengths comparable to the wakefield wavelength. These fragments act as a train of shorter pulses, driving bubbles in the normalized electron density ($n_e/n_1$) that roughly resemble those driven by a single short laser in the nonlinear regime. However, the variation and irregularity of these fragments during propagation degrade the wakefield and bubbles, significantly reducing the quality of the accelerated beam compared to the nonlinear regime. Moreover, while the overlap of the laser pulse fragments with the bubbles introduces DLA, it further degrades beam quality. Within the bubbles, particularly the two central ones ($-55 \,\si{\micro m} \le \xi \le -40 \,\si{\micro m}$), electrons are self-injected at the rear, drifting rightward as they are accelerated to energies close to 50 \si{MeV}. Oscillations of the accelerated electrons around the propagation axis are likely induced by the laser's transverse fields, possibly with an additional contribution from betatron motion \cite{Kiselev2004}. Along the down-ramp, as the density decreases, the wakefield wavelength and, consequently, the bubbles, become longer and eventually merge. Fig.~\ref{nelec2}(c), plotted halfway through the down-ramp, depicts this process. While the contour lines show a laser envelope similar to the one observed in the previous panel, the effect of reducing the plasma density can be seen in the normalized electron density $n_e/n_1$. The two central bubbles shown in Fig.~\ref{nelec2}(b) have merged in a single longer bubble in Fig.~\ref{nelec2}(c), loaded with almost continuously distributed electrons along its length. While the contour lines depict a laser envelope similar to the one observed in the previous panel, the effect of reducing the plasma density can be seen in the normalized electron density $n_e/n_1$. The two central bubbles shown in Fig.~\ref{nelec2}(b) have merged into a single, longer bubble in Fig.~\ref{nelec2}(c), filled with a nearly continuous distribution of electrons along its length. The two regions containing electrons exceeding 50 \si{\MeV} of kinetic energy (purple dots) were previously accelerated in individual bubbles before they merged. A key finding, unique to the optimal simulations, is the seamless transition of accelerated electrons between bubbles as they expand and merge, with the electrons maintaining their structure throughout their passage across the plasma. Fig.~\ref{nelec2}(d) shows the electrons with energies exceeding 8 \si{MeV} ejected from the gas target after exiting the down-ramp. The backtracked electrons presented in the previous panels were selected from this panel. Outside the gas target, the beam transversely expands due to the absence of focusing forces.

\section{\label{sec:results2} Optimization of a gas-density profile with two descending plateaus (Case 2)}

\begin{figure}
    \centering
    \includegraphics[scale=1]{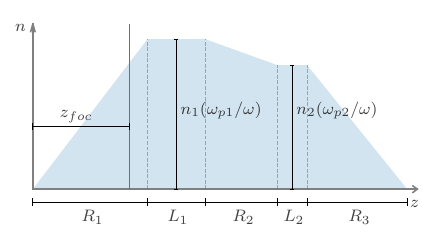}
    \caption{Gas-density profile with two descending plateaus (Case 2).}
    \label{gas2}
\end{figure}

If the long, fully loaded bubble observed in Fig.~\ref{nelec2}(c) could be maintained for a longer distance, the trapped electrons could potentially be accelerated to higher energies. This could be achieved by adding a second plateau, lower than the first one, at some point along the down-ramp. Such is the motivation for optimizing the gas-density profile shown in Fig.~\ref{gas2}, with two plateaus connected by a down-ramp, resembling two merged asymmetric trapezoidal profiles. The input parameters, increased from six to eight with the addition of a second plateau and down-ramp, are listed in Table \ref{tab:case2} along with their respective variation ranges. While the first plateau maintains the limits specified in Table \ref{tab:case1}, i.e., $0.15 \le \omega_{p,1}/\omega \le 0.40$, the second plateau's upper bound is constrained by the first, with $0 \le \omega_{p,2}/\omega \le \omega_{p,1}/\omega$.

\begin{table}[!t]
\caption{\label{tab:case2} Case 2: input parameters.}

\centering
\begin{tabular}{cccl}\hline

\multirow{2}{*}{\makecell[c]{Physical\\parameter}} & \multirow{2}{*}{\makecell[c]{Variation\\range}} & \multirow{2}{*}{\makecell[c]{Unit}} & \multirow{2}{*}{\makecell[c]{Description}}\\
&&\\
\hline
    $z_{foc}$ & $\phantom{00}5 \;,\, 200$&$\si{\um}$& Laser focal position\\[1ex]

    \multirow{3}{*}{\makecell[c]{$\omega_{p1}/\omega$}} & \multirow{3}{*}{\makecell[c]{$0.15 \;,\, 0.40$}} & \multirow{3}{*}{\makecell[c]{---}}  & \multirow{3}{*}{\makecell[l]{Plasma and laser\\frequencies ratio\\at the first plateau}}\\
    &&&\\
    &&&\\[1ex]
    \multirow{3}{*}{\makecell[c]{$\omega_{p2}/\omega$}} & \multirow{3}{*}{\makecell[c]{$\phantom{\omega_{p1}/}0 \;,\, \omega_{p1}/\omega$}} & \multirow{3}{*}{\makecell[c]{---}}  & \multirow{3}{*}{\makecell[l]{Plasma and laser\\frequencies ratio\\at the second plateau}}\\
    &&&\\
    &&&\\[1ex]

     $R_1$ & $ \phantom{0}10 \;,\, 120$&$ \si{\um}$ & First ramp length\\[1ex]
        $L_1$ & $\phantom{0}0 \;,\, 80$&$ \si{\um}$ & First plateau length \\[1ex]
          $R_2$ & $\phantom{0}10 \;,\, 120$&$ \si{\um}$ & Second ramp length\\[1ex]

        $L_{2}$ & $\phantom{0}0 \;,\, 60$&$ \si{\um}$ & Second plateau length \\[1ex]
      $R_{3}$ & $\phantom{00}0 \;,\, 120$&$ \si{\um}$ & Third ramp length\\
\hline
\end{tabular}
\end{table}

\begin{figure}
    \centering
\includegraphics{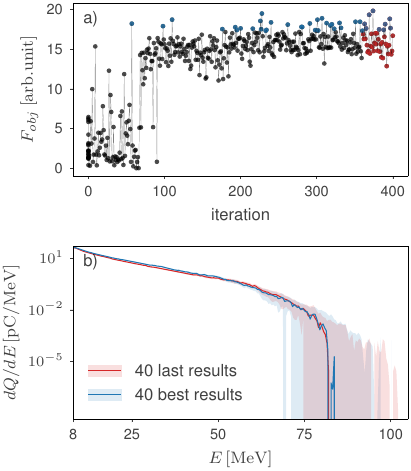}[!t]
    \caption{(a) $F_{obj}$ along the optimization of Case 2. Blue and red dots represent the 40 best and 40 last simulations, respectively. (b) Average energy spectra of the 40 best (blue line) and 40 last (red line) simulations, with shaded areas representing their 95\% confidence intervals.}
    \label{fig:caso2_iter}
\end{figure}

\begin{figure}[!t]
    \centering
    \includegraphics[scale=0.97]{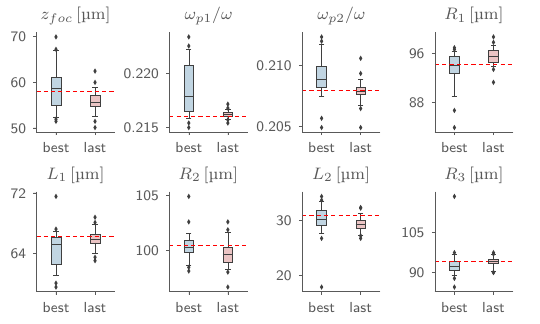}
    \caption{Input parameter distributions for the 40 best (blue boxplots) and 40 last (light-red boxplots) simulations of Case 2. The red dashed lines indicate the optimal values for each parameter.}  
    \label{fig:boxplots_caso2}
\end{figure}

Figure~\ref{fig:caso2_iter}(a) illustrates the evolution of the objective function over the iterations while optimizing Case 2. In the initial iterations, lower and highly variable values of $F_{obj}$ are observed as the algorithm explores the parameter space. However, after approximately 100 iterations, $F_{obj}$ values stabilize around $15.5\pm 1.8 \; \si{{a. u.}}$. This latter region, with reduced variability compared to the initial iterations, indicates the exploitation phase, where the optimization process refines its search around higher values of the objective function, demonstrating an effective convergence towards more optimal solutions. The optimal simulation (iteration 373) attained $F_{obj} \approx 19.8\, \si{{a. u.}}$, a value 66\% higher than the best result obtained for Case 1. The average spectra for the 40 best (blue dots) and 40 last (red dots) simulations, shown in Fig.~\ref{fig:caso2_iter}(b), do not exhibit any statistically significant differences. The distributions of their input parameters, as illustrated in Fig.~\ref{fig:boxplots_caso2}, are also compatible.

\begin{figure*}
    \centering
\includegraphics{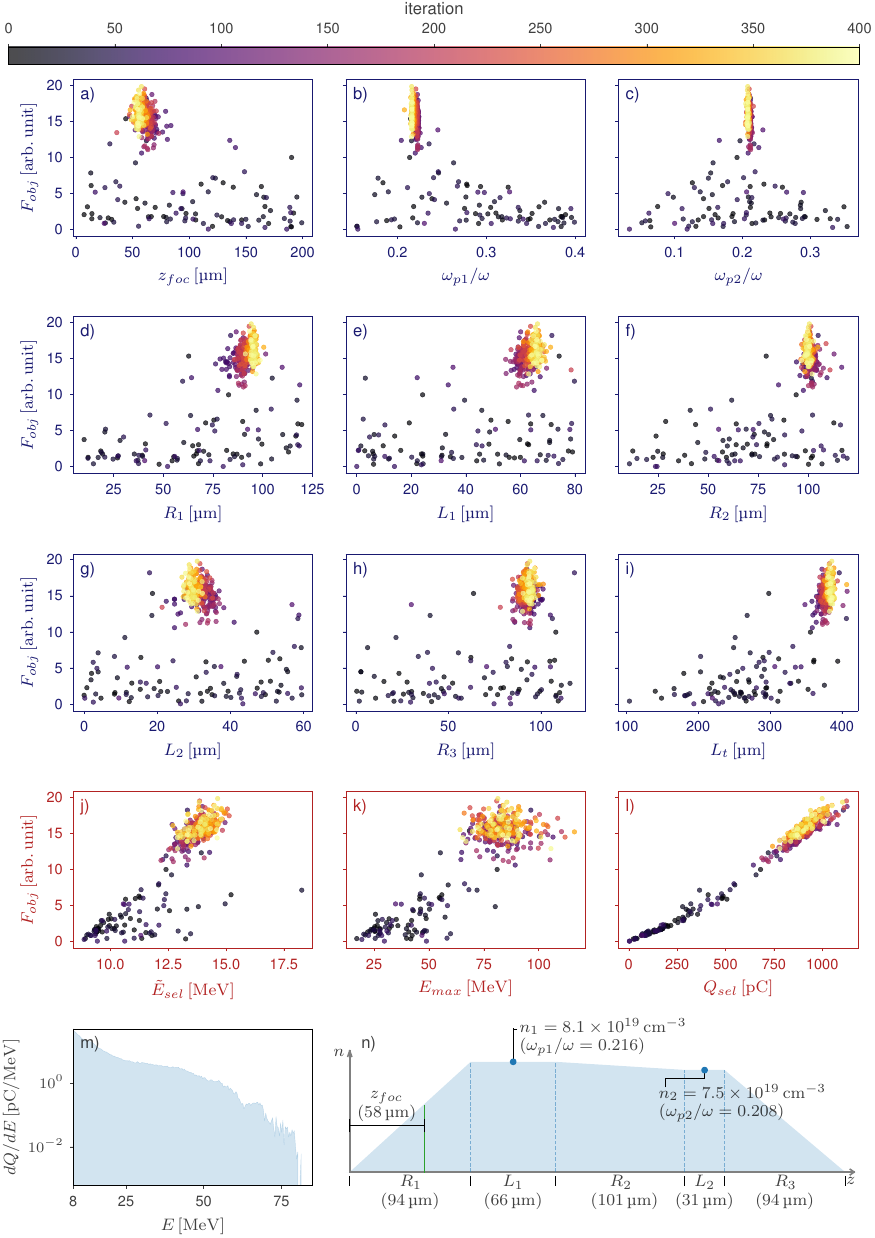}
    \caption{Bayesian optimization of Case 2, with a color scale indicating the algorithm iteration. Input parameters: $F_{obj}$ versus the (a) laser focal position, plasma-to-laser frequency ratio in the (b) first and (c) second plateaus, lengths of the (d) first ramp, (e) first plateau, (f) central ramp, (g) second plateau, (h) last ramp and (i) total profile. Output parameters: $F_{obj}$ versus the (j) median energy (above 8 \si{MeV}) $\tilde{E}_{sel}$, (k) maximum energy, and (l) selected charge. Case 2 optimal: (m) energy spectrum and (n) gas-density profile, with optimal input parameters.}
    \label{resul_caso3}
\end{figure*}

Figures~\ref{resul_caso3}(a)-(i) show scatter plots of the objective function against the input parameters listed in Table \ref{tab:case2}, with the total length $L_{total}$ obtained by adding the length of all ramps and plateaus. In all panels, narrow clusters of light-colored dots, located at higher values of $F_{obj}$, indicate the optimal values of the input parameters. Figures~\ref{resul_caso3}(j)-(l) present the beam output parameters accordingly. After convergence, Fig.~\ref{resul_caso3}(j) shows median energies (above 8 \si{MeV}) between 13 \si{\MeV} and 16 \si{\MeV}, while Fig.~\ref{resul_caso3}(k) depicts maximum energies between 75 \si{\MeV} and 120 \si{\MeV}. Additionally, Fig.~\ref{resul_caso3}(l) highlights selected charges from 800 \si{\pico \coulomb} to 1200 \si{\pico \coulomb} after convergence. In the optimal simulation of Case 2 ($F_{obj} = 19.8$), a median energy $\tilde{E}_{sel} = 14.6$ \si{\MeV}, maximum energy $E_{max} = 81$ \si{\MeV}, and selected charge $Q_{sel} = 999$ \si{\pico \coulomb} were attained. Despite the lower median energy compared to Case 1, the 67\% increase in the optimal charge in Case 2 led to a higher value for the objective function. Additionally, the longer total length $L_{total}$ also benefited Case 2. To conclude Fig.~\ref{resul_caso3}, panels (m) and (n) display the optimal beam energy spectrum and gas-density profile, respectively. 

\begin{figure*}[!t]
    \centering
\includegraphics[scale=1]{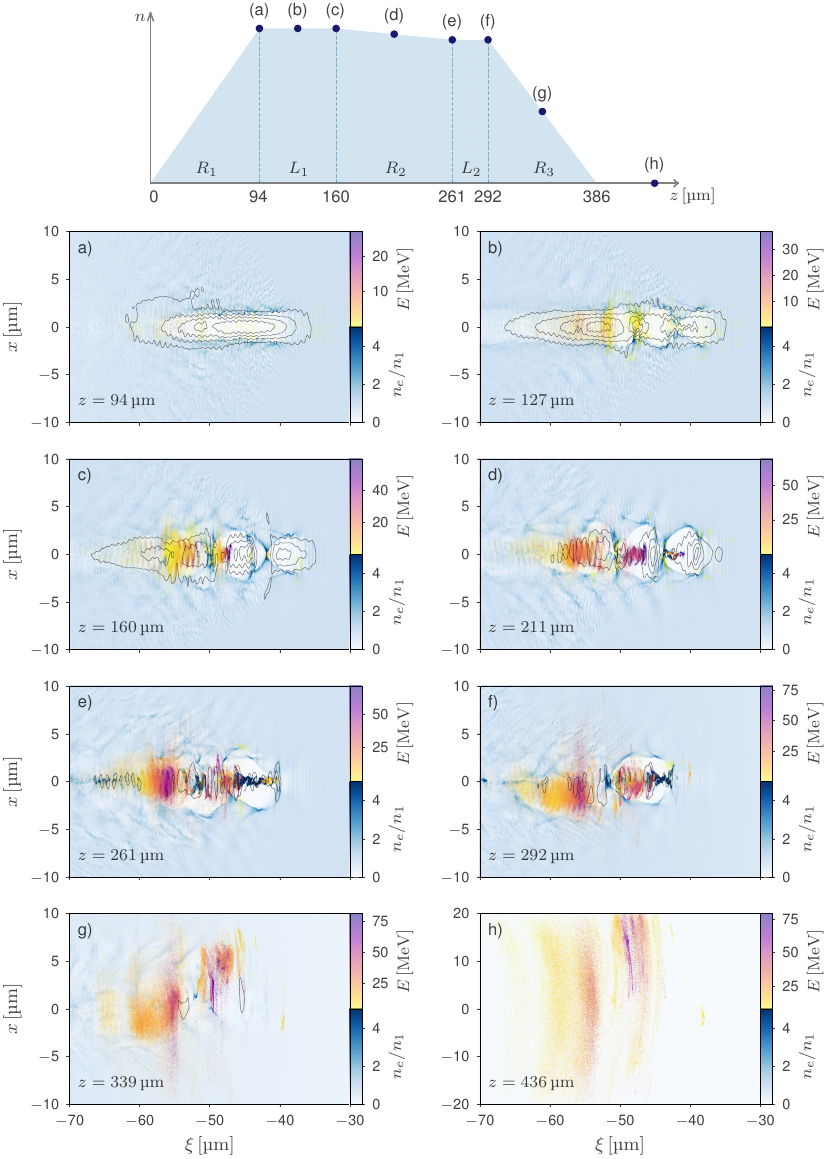}
        \caption{Spatial evolution of the optimal simulation, Case 2. The upper panel shows the propagation distances at which panels (a) to (h) were plotted. In these panels, the plasma electron density, normalized by the plateau density, is shown using a blue gradient scale. The backtracked electrons, which will attain energies exceeding 8 \si{MeV} after ejection, are depicted as colored dots with a yellow-purple color scale representing their energies at the current propagation distance. Additionally, contour lines for $a_0$ ranging from 1 (outermost) to 4 (innermost) were added to show the laser envelope at these distances.} 
    \label{nelec3a}
\end{figure*}

Figure~\ref{nelec3a} displays the laser and plasma dynamics for the optimal simulation of Case 2. Panels (a) to (c), plotted along the first plateau ($L_1$), exhibit behavior qualitatively comparable to Figs.~\ref{nelec2}(a)-(b), showing the same region of the optimal profile in Case 1. However, additional contour lines of $a_0$ reveal higher-intensity laser fragments after the self-modulation in Case 2. Figures~\ref{nelec3a}(d)-(e) show the laser descending the first density down-ramp ($R_2$). In panel~(d), plotted in the middle of the descent, the plasma still contains three bubbles. Additionally, the more intense laser fragments are capable of accelerating electrons to higher energies, above 50 \si{MeV}, as indicated by the purple dots. Panel~(e) shows that, at the end of the ramp and the beginning of the second plateau ($L_2$), the second and third bubbles, counted in the decreasing direction of $\xi$, are now merging into a single longer one. A notable difference observed in the optimal simulation of Case 2, compared to Case 1, is that the presence of a longer and less steep down-ramp allowed for a strong self-injection of electrons within the merging bubbles. Regarding the laser along the down-ramp $R_2$, while panel~(d) still shows intense fragments, with lengths comparable to the bubbles, panel~(e) displays a larger number of shorter fragments, with lengths now proportional to the laser wavelength. Although most of these fragments have lower intensities ($a_0 \gtrsim 1$), fragments with $a_0 \gtrsim 3$ are still capable of driving the observed bubbles up to the end of second plateau ($L_2$), as shown in Fig.~\ref{nelec3a}(f). The eccentricity observed in this panel -- the second bubble is slightly off-axis -- intensifies in panels (g) and (f), plotted halfway down the final down-ramp ($R_3$) and outside the gaseous target, respectively. As a consequence, the accelerated beam, shown in the latter panel, exhibits significant transverse spreading.

\section{\label{sec:results3}Optimization of a generalized two-plateau gas-density profile (Case 3)}

\begin{figure}
    \centering
    \includegraphics{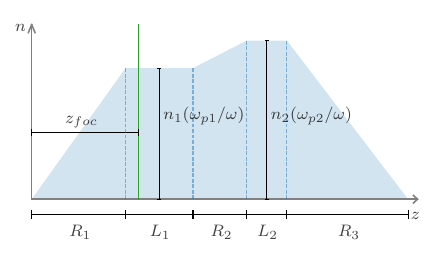}
    \caption{Generalized two-plateau gas-density profile (Case 3).}
    \label{gas3}
\end{figure}

While Case 2 constrains the second density plateau to be always lower than the first one ($0 \le \omega_{p2}/\omega \le \omega_{p1}/\omega$), Case 3, illustrated in Fig.~\ref{gas3}, adopts a generalized two-plateau gas-density profile. In this case, the second plateau is allowed to vary within a fixed range ($0 \le \omega_{p2}/\omega \le 1$, as shown in Table~\ref{tab:case3}), eliminating the variable, $\omega_{p1}/\omega$-dependent upper limit previously adopted.

\begin{figure}
    \centering
\includegraphics[scale=1]{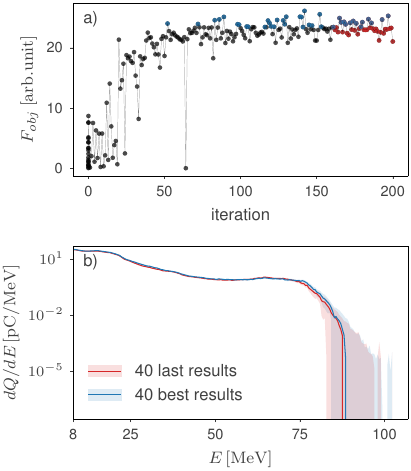}
    \caption{(a) $F_{obj}$ along the optimization of Case 3. Blue and red dots represent the 40 best and 40 last simulations, respectively. (b) Average energy spectra of the 40 best (blue line) and 40 last (red line) simulations, with shaded areas representing their 95\% confidence intervals.}
    \label{fig:caso3_iter}
\end{figure}

\begin{figure}
    \centering
    \includegraphics[scale=0.97]{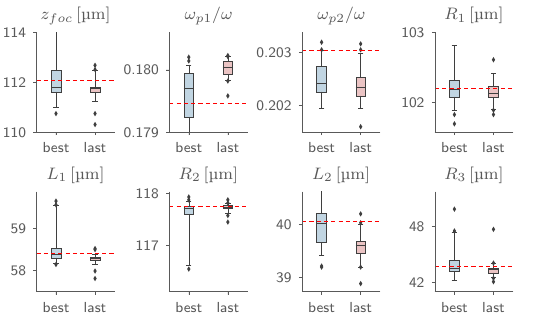}
    \caption{Input parameter distributions for the 40 best (blue boxplots) and 40 last (light-red boxplots) simulations of Case 3. The red dashed lines indicate the optimal values for each parameter.}  
    \label{fig:boxplots_caso3}
\end{figure}

Figure~\ref{fig:caso3_iter}(a) shows convergence around $F_{\text{obj}} \approx 23 \pm 1.2$ u.a. after approximately 70 iterations, with the optimal simulation (at iteration 142) achieving $F_{\text{obj}} = 26.1$ u.a. In contrast to previous cases, optimization in Case 3 stopped at 200 iterations. Given the achieved convergence and considering the behavior of previous cases, this early termination reduced computational costs without compromising the analysis. As shown in Fig.~\ref{fig:caso3_iter}(b), the average spectra of the 40 best and 40 last simulations are statistically equivalent. In the previous two cases, the typical logarithmic decay shows a negative slope, indicating an exponential decrease in charge distribution $dQ/dE$ with increasing energy, However, in Case 3, this decay flattens out over certain energy ranges, forming distinct plateaus. There is an almost constant charge distribution between approximately 40 and 60 \si{MeV}, followed by a higher plateau between approximately 60 and 80 \si{MeV}. Figure \ref{fig:boxplots_caso3} exhibits the input parameter distributions of the 40 best and 40 last simulations.

\begin{table}
\caption{\label{tab:case3} Case 3: input parameters.}
\centering
\begin{tabular}{@{}cccl}
\hline
\multirow{2}{*}{\makecell[c]{Physical\\parameter}} & \multirow{2}{*}{\makecell[c]{Variation\\range}} & \multirow{2}{*}{\makecell[c]{Unit}} & \multirow{2}{*}{\makecell[c]{Description}}\\
&&\\
\hline

    $z_{foc}$ & $\phantom{00}5 \;,\, 200$ & $ \si{\um}$& Laser focal position\\[1.0ex]
    \multirow{3}{*}{\makecell[c]{$\omega_{p1}/\omega$}} & \multirow{3}{*}{\makecell[c]{$0.15 \;,\, 0.40$}} & \multirow{3}{*}{\makecell[c]{---}}  & \multirow{3}{*}{\makecell[l]{Plasma and laser\\frequencies ratio at\\the first plateau}}\\
    &&&\\
    &&&\\[1.0ex]
    \multirow{3}{*}{\makecell[c]{$\omega_{p2}/\omega$}} & \multirow{3}{*}{\makecell[c]{$0 \;,\, 1$}} & \multirow{3}{*}{\makecell[c]{---}}  & \multirow{3}{*}{\makecell[l]{Plasma and laser\\frequencies ratio at\\the second plateau}}\\
    &&&\\
    &&&\\[1.0ex]

     $R_1$ & $ \phantom{0}10 \;,\, 120$&$ \ \si{\um}$ & First ramp length\\[1.0ex]
        $L_1$ & $\phantom{0}0 \;,\, 80$&$ \ \si{\um}$ & First plateau length \\[1.0ex]
          $R_2$ & $\phantom{0}10 \;,\, 120$&$ \ \si{\um}$ & Second ramp length\\[1.0ex]

        $L_{2}$ & $\phantom{0}0 \;,\, 60$&$ \ \si{\um}$ & Second plateau length \\[1.0ex]
      $R_{3}$ & $\phantom{00}0 \;,\, 120$&$ \ \si{\um}$ & \makecell[l]{Third ramp horizontal \\ length}\\
\hline
\end{tabular}
\end{table}

\begin{figure*}
    \centering
\includegraphics{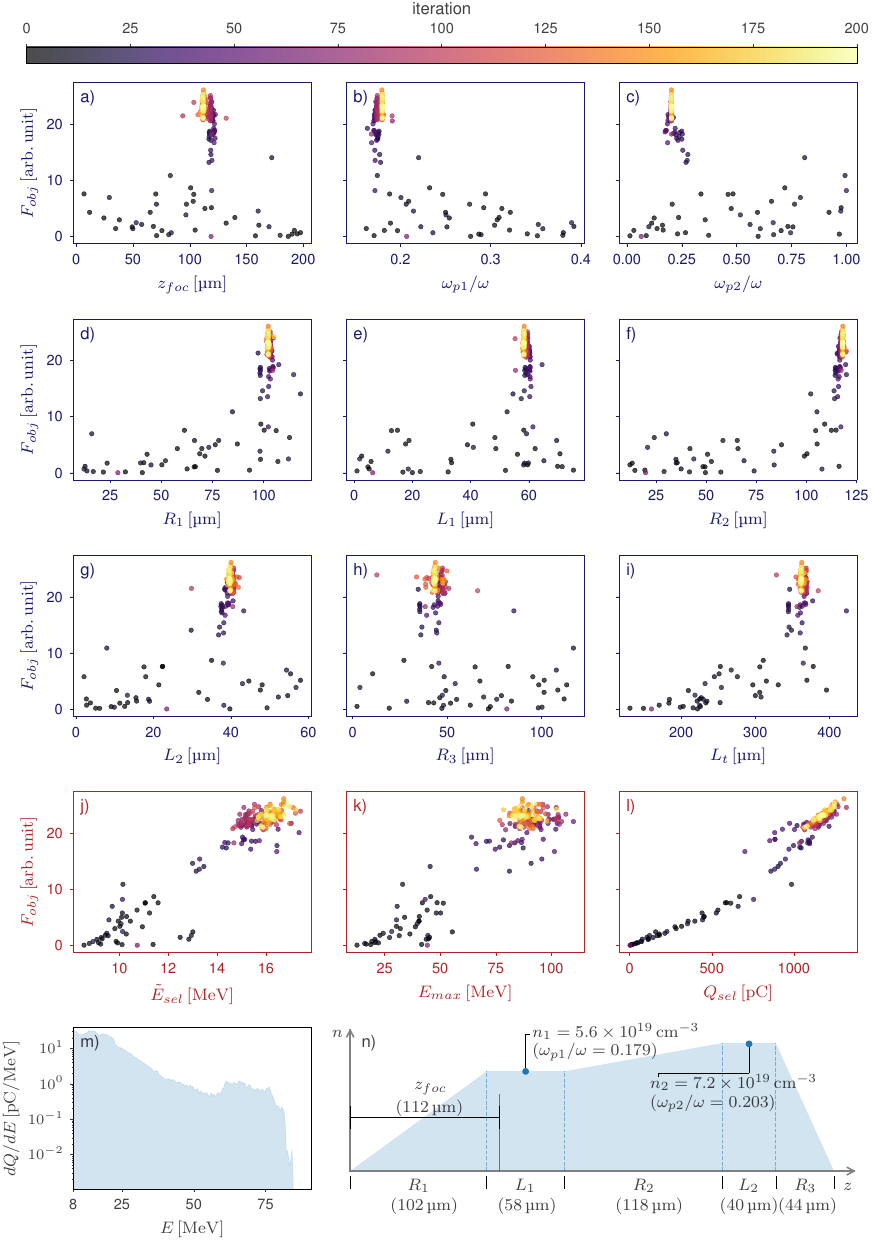}
    \caption{Bayesian optimization of Case 3, with a color scale indicating the algorithm iteration. Input parameters: $F_{obj}$ versus the (a) laser focal position, plasma-to-laser frequency ratio in the (b) first and (c) second plateaus, lengths of the (d) first ramp, (e) first plateau, (f) central ramp, (g) second plateau, (h) last ramp and (i) total profile. Output parameters: $F_{obj}$ versus the (j) median energy (above 8 \si{MeV}) $\tilde{E}_{sel}$, (k) maximum energy, and (l) selected charge. Case 2 optimal: (m) energy spectrum and (n) gas-density profile, with optimal input parameters.}
    \label{resul_caso4}
\end{figure*}

Figures~\ref{resul_caso4}(a)-(i) show scatter plots of the objective function against  the input parameters from Table \ref{tab:case3}, plus the total target length ($L_{total}$). By removing the constraint of a lower density in the second plateau from Case 2, the optimization in Case 3 resulted in a distinct configuration. The central ramp ($R_2$) transitioned from a down-ramp in Case 2 to a longer and steeper up-ramp in Case 3. Despite this, the optimal total length in Case 3 (362 \si{\micro m}) is shorter than in Case 2 (386 \si{\micro m}). Regarding the output beams, Fig.~\ref{resul_caso4}(j) shows median energies (above 8 \si{MeV}) converging around 16 \si{\MeV}, with the optimal simulation achieving $\tilde{E}_{sel}=16.7$ \si{\MeV}. Figure~\ref{resul_caso4}(k) depicts maximum energies ranging from 80 \si{\MeV} to 110 \si{\MeV} after convergence, with the optimal simulation reaching a maximum energy of 87 \si{\MeV}. In Fig.~\ref{resul_caso4}(l), total charges (above 8 \si{\MeV}) varied from 1050 \si{\pico \coulomb} to 1310 \si{\pico \coulomb} after convergence, with $Q_{sel} = 1305 \, \si{\pico \coulomb}$ for the optimal simulation (an approximate 30\% increase compared to Case 2). The optimal electron beam spectrum and the corresponding gas-density profile are presented in (m) and (n), respectively.

\begin{figure*}[!t]
    \centering
\includegraphics[scale=1]{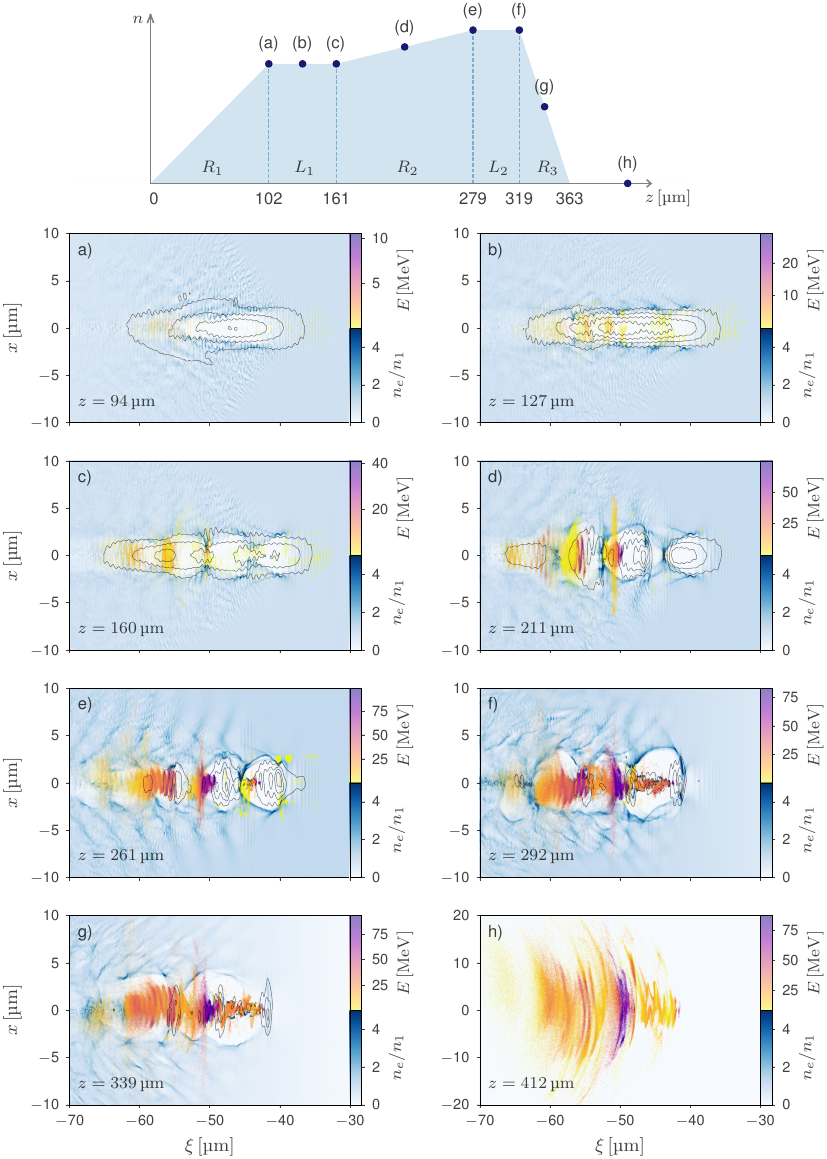}
        \caption{Spatial evolution of the optimal simulation, Case 3. The upper panel shows the propagation distances at which panels (a) to (d) were plotted. In these panels, the plasma electron density, normalized by the plateau density, is shown using a blue gradient scale. The backtracked electrons, which will attain energies exceeding 8 \si{MeV} after ejection, are depicted as colored dots with a yellow-purple color scale representing their energies at the current propagation distance. Additionally, contour lines for $a_0$ ranging from 1 (outermost) to 4 (innermost) were added to show the laser envelope at these distances.}
    \label{nelec4a}
\end{figure*}

The laser and plasma dynamics for the optimal simulation of Case 3 are depicted in Fig.~\ref{nelec4a}. Panels (a) to (c) illustrate the system along the first plateau ($L_1$). Compared to Case 2, the laser self-modulates later due to the longer focal position in Case 3 ($z_{foc} = 112 \,\si{\micro m}$, as opposed to $58 \,\si{\micro m}$ in Case 2). At the end of $L_1$, shown in panel (c), two continuous external contour lines ($a_0 = 1$ and $a_0 = 2$) still envelop four peaks exceeding $a_0 = 3$. Despite this, the plasma density shows well-formed bubbles, and a small fraction of backtracked electrons has already reached energies above 75 \si{MeV}. As depicted in panels (d) and (e), the self-modulation intensifies as the laser propagates along the up-ramp ($R_2$) and reaches the second plateau ($L_2$). During this process, the bubble structure of the plasma electrons is preserved, and a significant injection of backtracked electrons is observed in this region. Moreover, because the wakefield amplitude scales with the increased density, a greater fraction of these electrons is accelerated to higher energies. At the end of the second plateau, as depicted in panel (f), the system resembles a laser wakefield accelerator (LWFA) operating in the nonlinear regime, with a train of short and intense laser fragments driving well-defined bubbles filled with high-energy electrons. Panel (g) shows that halfway down the final ramp ($R_3$), which is approximately 50\% shorter than the corresponding ramp in Case 2, the main laser fragments and the accelerating structure remain well-defined. As observed in the optimal simulations of previous cases, there is a seamless transition of accelerated electrons between bubbles as they expand and merge, facilitated due to the decreasing density. The final panel of Fig.~\ref{nelec4a}, panel (h), depicts the accelerated beam after ejection from the plasma.

\section{\label{sec:comparison}Comparison of the three optimized gas-density profiles}

\begin{figure}
    \centering
    \includegraphics[scale=0.95]{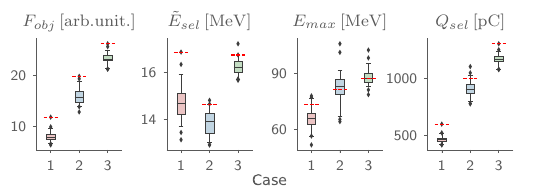}
    \caption{Output parameter distributions from the 40 last simulations of Cases 1, 2, and 3 (optimal values are marked with red dashed lines).}
    \label{fig:boxplots_40_last}
\end{figure}

Table \ref{tab:optimal_results} presents the optimal input and output parameters for Cases 1, 2, and 3, illustrating the differences across these scenarios. Case 1 exhibits the least favorable performance, with an objective function $F_{obj}$ of 11.9 \si{a.u.}, a median selected energy $\tilde{E}_{sel}$ of 16.9 \si{MeV}, and a selected charge $Q_{sel}$ of approximately 600 \si{pC}. In contrast, Case 2, despite having a lower median energy, $\tilde{E}_{sel} = 14.6 \; \si{MeV}$, achieves a significantly higher objective function, $F_{obj} = 19.8 \; \si{a.u.}$. This improvement is attributed to a substantial 67\% increase in the charge, with $Q_{sel} \simeq 1000 \; \si{pC}$. Case 3 exhibits the best performance among the three cases, with $F_{obj} = 26.1 \; \si{a.u.}$. Although the median energy $\tilde{E}_{sel} = 16.7 \; \si{MeV}$ in Case 3 is higher than in Case 2, it did not reach the 16.9 \si{MeV} attained in Case 1. However, a 30\% increase in the charge, $Q_{sel} \simeq 1300 \; \si{pC}$, ensured significant improvement of Case 3 over Case 2. Fig.~\ref{fig:boxplots_40_last} corroborates the consistence of these findings by comparing the distributions of the objective function and the output parameters from the last 40 simulations of each case, with the optimal values marked by red dashed lines. 

\begin{figure}
    \centering
    \includegraphics[scale=1]{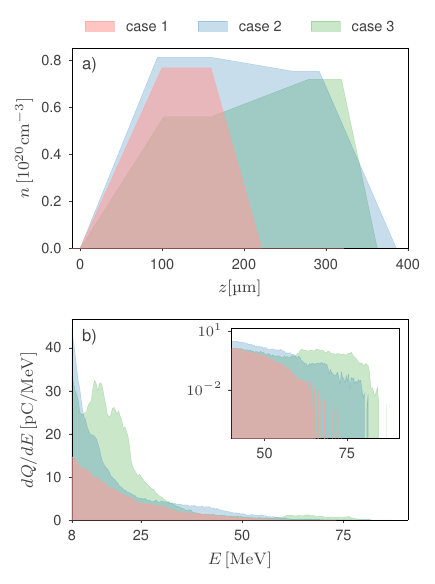}
    \caption{Optimal (a) profiles and (b) energy spectra, with higher energies displayed in a log-scale inset.}
    \label{fig:optimal_results}
 \end{figure}

Fig.~\ref{fig:optimal_results}(a) displays the gas-density profiles of the optimal gas configurations for the three cases. The shorter total length of Case 1 contributes for its lower performance, as the electrons are accelerated over a shorter distance. Up to the end of the first plateau ($L_1$), the profile in Case 2, despite its higher density, is comparable to that of Case 1. However, in Case 2, $L_1$ is followed by a longer and less steep down-ramp ($R_2$) connecting the first plateau to a second one ($L_2$), which is shorter and has a slightly lower density. The profile then concludes with a final down-ramp ($R_3$), with length and declivity similar to those in Case 1. In Case 2, the addition of a second, lower-density plateau starting from the down-ramp aimed at increasing the acceleration length while maintaining the system configuration established along the down-ramp. Although this intended goal was not fully achieved, as evidenced by the lower median energy in Case 2 compared to Case 1, the proposed profile significantly enhanced the self-injection of plasma electrons. This enhancement was particularly effective in the down-ramps, where decreasing density allows the bubbles to elongate and capture more electrons at their rear. This mechanism, depicted in Fig.~\ref{nelec3a}, was identified by the algorithm as the optimal strategy to maximize the objective function by increasing the beam charge. In Case 3, the constraint of having the second plateau with a lower density than the first one was removed. This change allowed the algorithm to find a distinct optimal profile, with an up-ramp connecting a lower-density plateau to a higher-density one. This configuration produced a 30\% increase in charge compared to the optimal result of Case 2, and yielded a spectrum favorable for producing bremsstrahlung photons with energies capable of triggering the desired photoactivation reaction.

Fig.~\ref{fig:optimal_results}(b) depicts the optimal energy spectra for the three cases. Case 3 clearly outperforms Cases 1 and 2, deviating from the typical exponential decay commonly observed in electron beam spectra accelerated via SM-LWFA. The charge peak within the 8-25 \si{MeV} range in Case 3 is particularly interesting, as it might increase the yield of photons with optimal energies for the $^{\text{100}}$Mo($\gamma$,n)$^{99}$Mo reaction, enhancing the production of $^{99}$Mo. Additionally, the optimal spectrum of Case 3 exhibits a distinct high-charge plateau at energies exceeding $\sim$50 \si{MeV}, as highlighted in the inset of Fig.~\ref{fig:optimal_results}. This can be attributed to the up-ramp placed after the wakefield and bubbles were established along the first plateau \cite{Aniculaesei2019}.

\begin{figure}[!t]
\centering
\includegraphics[scale=1]{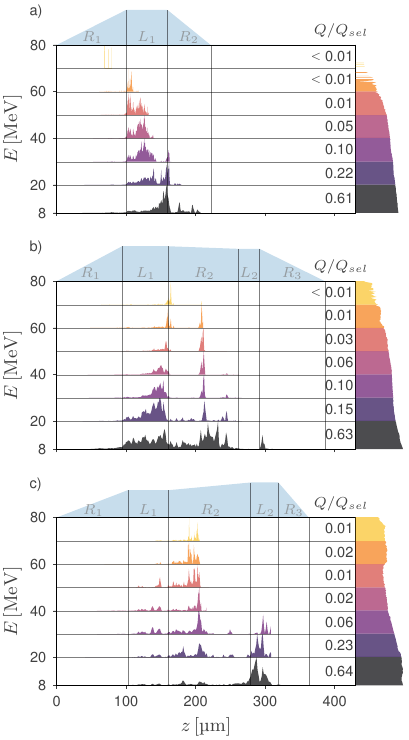}
\caption{Propagation distances at which the backtracked electrons (comprising the accelerated beams) originated along the optimal profiles.}
\label{fig:histograms_z}
\end{figure}

To visualize how the ramps and plateaus affect the accelerated beams, Figs.~\ref{fig:histograms_z}(a)--(c) display histograms of the propagation distances at which the backtracked electrons (comprising the accelerated beams) originated along the optimal profiles. These histograms are stratified and color-coded by energy ranges that cover the corresponding optimal spectra, displayed on the right side of each panel. Additionally, the fraction of charge contained within each energy range, denoted as $Q / Q_{sel}$, is also displayed. Fig.~\ref{fig:histograms_z}(a) shows that, in Case 1, the majority of electrons of the accelerated beam originated along the plateau $L_1$, with exceptions being a small fraction of high-energy particles from the up-ramp $R_1$, and a low-energy fraction from the down-ramp $R_2$. Additionally, the histograms indicate that backtracked electrons with higher energies originated from earlier positions of $L_1$, likely because they were accelerated over longer distances. In Case 2, Fig.~\ref{fig:histograms_z}(b) shows that while most of the backtracked electrons originated along the first plateau $L_1$, an appreciable fraction also originated halfway down the ramp $R_2$, across most energy ranges. Interestingly, in contrast to Case 1, electrons originated from later positions in $L_1$ achieved higher energies in the accelerated beam. In Case 3, Fig.~\ref{fig:histograms_z}(c) illustrates that, despite the contribution from the plateau $L_1$, the up-ramp $R_2$ plays a crucial role in forming the accelerated beam, contributing appreciable charge across all energy ranges. The second plateau $L_2$ also contributes a large fraction of the beam charge, mostly in the lower energy ranges. Nevertheless, since these ranges coincide with the maximum cross-section for the $^{99}$Mo($\gamma$, $n$)$^{100}$Mo reaction, these particles may play a relevant role in the production of $^{99}$Mo. The observed results arise from an intricate overlap of two accelerating mechanisms, LWFA and DLA. Their relative contributions depend on the coupled laser-plasma dynamics, influenced by variations in the gas-density profile and laser self-modulation. A deeper analysis of the contributions of each of the accelerating mechanisms for the three optimal cases will be presented in an upcoming work.

To conclude the comparison, preliminary estimates for the production of $^{\text{99}}$Mo were obtained for the three cases, using Monte Carlo simulations executed with the TOPAS~ \cite{Perl2012,Faddegon2020} code. For each optimal beam, $10^8$ electrons were randomly sampled from their FBPIC phase space, taking into account the macroparticle weights. These electrons were then used in TOPAS as the radiation source to irradiate a cylindrical target made of natural molybdenum, measuring 5 \si{\cm} in length and 10 \si{\cm} in radius, positioned 10 \si{\cm} from the source in vacuum. The front face of the target features a 4.5 \si{\mm}-thick layer of tantalum to convert the impinging electrons into photons through bremsstrahlung. The \textit{g4em-standard\_opt4} and \textit{g4em-extra} physics lists were used, and an \textit{OriginCount} volume scorer was set to count the $^{99}$Mo yield within the target.

\begin{table}
    \caption{\label{tab:optimal_results} Optimal input and output parameters.}
    \centering
    \begin{tabular}{ccccc}\hline
        \multirow{2}{*}{Parameter}            & \multirow{2}{*}{Unit}              & \multirow{2}{*}{Case 1} & \multirow{2}{*}{Case 2} & \multirow{2}{*}{Case 3} \\
        &&&&\\
        \hline
        $z_{foc}$            &\si{\um}	         & 113.4  & 57.9   & 112.1 \\[0.70ex]
        $\omega_{p1}/\omega$ &---                & 0.210  & 0.216  & 0.179 \\[0.70ex]
        $\omega_{p2}/\omega$ &---                & ---    & 0.208  & 0.203 \\[0.70ex]
        $R_1$                & \si{\um}    	     & 99.8   & 94.2   & 102.3 \\[0.70ex]
        $L_1$                & \si{\um}          & 59.8   & 66.3   & 58.4  \\[0.70ex]
        $R_2$                & \si{\um}    	     & 62.1   & 100.5  & 118.2 \\[0.70ex]
        $L_2$                &\si{\um}    	     & ---    & 30.9   & 40.1  \\[0.70ex]
        $R_3$                & \si{\um}    	     & ---    & 94.1   & 43.7  \\[0.70ex]
        \\[-1.5ex]
        $F_{obj}$           & \si{{arb. unit}}   & 11.9   & 19.8   & 26.1  \\[0.70ex]
        $\tilde{E}_{sel}$   &\si{\MeV}           & 16.9   & 14.6   & 16.7  \\ [0.70ex]
        $E_{max}$           &\si{\MeV}           & 73     & 81     & 87    \\[0.70ex]
        $Q_{sel}$           & \si{\pico\coulomb} & 601    & 999    & 1305 \\
        \hline
    \end{tabular}
\end{table}

Table~\ref{tab:99Mo_per_shot} displays the number of $^{99}$Mo isotopes ($N$) produced by a single shot from each optimal beam. Despite the qualitative variations in the optimal energy spectra shown in Fig.~\ref{fig:optimal_results}(b), Table~\ref{tab:99Mo_per_shot} reveals a strong scaling between the $^{99}$Mo yield and the beam charge. This is evident from the normalized quantities $N/N_1$ and $Q/Q_{sel,1}$, with $N_1$ and $Q_{sel,1}$ representing Case 1's $^{99}$Mo yield and beam charge, respectively.

\begin{figure}
\centering
\includegraphics[scale=1]{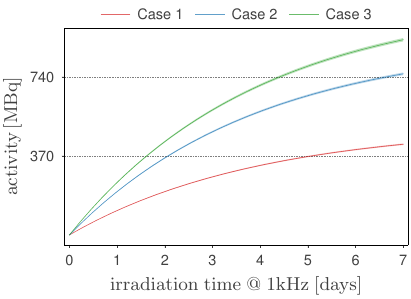}
\caption{$^{\text{99}}$Mo activity as a function of irradiation time, assuming a repetition rate of 1 \si{\kHz}.}
\label{fig:activity}
\end{figure}

The administered activity of a given radiopharmaceutical depends on the intended medical procedure. For $^\text{99m}$Tc, the daughter isotope of $^{99}$Mo, a typical activity of 370 \si{\mega\becquerel} per patient is often used, although US standards suggest doubling this to 740 \si{\mega\becquerel} \cite{Vieira2021,Drozdovitch2015}. Given the longer half-life of $^{99}$Mo (66 hours) relative to $^\text{99m}$Tc (6 hours), the activity of the daughter isotope, though slightly lower, is similar to that of its parent. Hence, the $^{99}$Mo activities illustrated in Fig.~\ref{fig:activity}, derived from the yields listed in Table~\ref{tab:99Mo_per_shot} assuming an SM-LWFA operating at 1 kHz, can also be used as estimates for those of $^\text{99m}$Tc. From Fig.~\ref{fig:activity}, after 7 days of irradiation, Case 1 barely surpasses 370 \si{\mega\becquerel}, while Case 2 reaches the 740 \si{\mega\becquerel} threshold. Case 3 shows the best performance, exceeding this threshold in approximately 4.5 days. Although these results are not yet practical for clinical applications, they represent a significant improvement over those reported in a previous computational study using LWFA-accelerated electrons \cite{Vieira2021}. Moreover, employing multiple laser systems could substantially reduce the irradiation time. For instance, working with five laser systems in parallel and using Case 3 optimal parameters, the 740 \si{\mega\becquerel} threshold could be reached in approximately 13.7 hours of irradiation.

\begin{table}
\caption{\label{tab:99Mo_per_shot}Number of $^{99}$Mo isotopes (including the standard deviation) produced by a single shot of each optimal beam.}
\centering
\renewcommand{\arraystretch}{1.15} 
\begin{tabular}{ccccc}\hline
Case & \makecell{$Q_{sel}$\\ $\mathrm{(pC)}$} & \makecell{$N \times 10^6$ \\(\text{atoms of $^{99}$Mo})} & \textbf{$Q_{sel}/Q_{sel,1}$} & \textbf{$N/N_1$} \\
\hline
1 & 601 & $0.513 \; (\pm 0.004$) & 1 & 1 \\[1.0ex]
2 & 999 & $0.911 \; (\pm 0.008$) & 1.66 & 1.78 \\[1.0ex]
3 & 1305 & $1.103 \; (\pm 0.009$) & 2.17 & 2.15 \\
\hline
\end{tabular}
\end{table}

\section{\label{sec:conclusions}Conclusions and discussion}

In this work, Bayesian optimization was applied within PIC simulations to maximize the energy and charge of electron beams produced by laser wakefield acceleration in the self-modulated regime. Optimization of three distinct profiles produced electron beams with median energies between 14 to 17 \si{MeV} and charges ranging from 600 to 1300 \si{\pico\coulomb}, considering electrons with energies above 8 \si{MeV}. These beams can generate bremsstrahlung photons capable of producing $^{99}$Mo through the $^{99}$Mo($\gamma$, $n$)$^{100}$Mo reaction route in a tantalum-molybdenum target. Monte Carlo simulations estimated preliminary $^{99}$Mo yields for each configuration, enabling calculation of irradiation times to achieve activity levels of 370 and 740 \si{\mega\becquerel}, comparable to the clinical activity of its daughter isotope, $^\text{99m}$Tc, used in medical procedures.

For the optimal gas-density profiles, the laser-plasma dynamics along the ramps and plateaus was presented and discussed. Additionally, electrons comprising the final accelerated beams were backtracked to assess their energy gains throughout propagation. Energy-binned histograms were plotted for the propagation distances at which the backtracked electrons were trapped/injected, providing insights into how the ramps and plateaus influence the final energy and charge distributions in the accelerated beams. In the self-modulated regime, both acceleration mechanisms, LWFA and DLA, coexists. As the laser pulse undergoes self-focusing and self-modulation, the formation of short, intense laser fragments occurs. These fragments are capable of driving bubbles similar to those observed in the nonlinear regime of LWFA. However, due to the presence of multiple fragments, DLA is caused by the overlap of these fragments with the electrons undergoing LWFA acceleration within the bubbles. A future study will be conducted to provide a detailed evaluation of the respective roles played by LWFA and DLA in achieving these optimal beams.

Tailored profiles have been investigated for various applications, including plasma beam dumps~\cite{Jakobsson2019} and proton-driven wakefield acceleration in Project AWAKE~\cite{Adli2018}. Although experimentally challenging, the implementation of such profiles might be feasible using de Laval nozzles or one-sided shock nozzles~\cite{Rovige2020} to generate appropriate gas jets. Alternatively, when using capillaries, temperature gradients can create density ramps, while sections held at constant temperatures can form plateaus of varying densities. This method was successfully applied to tailor the 10-meter-long rubidium plasma in Project AWAKE~\cite{Plyushchev2017}. Additionally, while the presented simulations utilized only hydrogen, future computational studies will explore and optimize gas mixtures, such as hydrogen with nitrogen, to further enhance the energy and charge of electron beams produced by laser wakefield accelerators operating in the self-modulated regime.

While Bayesian optimization effectively maximized electron beam charge and energy, this strategy may not be ideal for $^{99}$Mo photoactivation. The chosen objective function, prioritizing high-energy particles (charge $\times$ energy), yielded spectra exceeding 70 \si{MeV}. Though these electrons can generate suitable photons (8-20 \si{MeV}) for $^{99}$Mo photoactivation, they can also produce higher-energy photons, capable of inducing unwanted photonuclear reactions, leading to neutron and secondary particle emissions \cite{Martin2017}. This risk must be assessed before using beams with high-energy electrons for isotope production via photoactivation.

The $^{99}$Mo yields presented in this work are preliminary, non-optimal results, intended primarily to provide an expected order of magnitude and to allow for relative comparisons between the cases presented. In a previous study~\cite{Broder2022}, TOPAS demonstrated mixed performance in modeling radioisotope production compared to GEANT4~\cite{Agostinelli2003}. Key issues include limitations in physics lists, especially at lower energies for light target nuclei, and high sensitivity to the target geometry and placement. Hence, further investigation is required to attain more reliable results. While the current $^{99}$Mo yields fall short of clinical practicality, they represent a significant advancement over a prior SM-LWFA-based approach~\cite{Vieira2021}. Furthermore, multiple high-repetition-rate, few-terawatt laser systems could be employed in parallel for achieving clinically relevant activities (370-740 \si{\mega\becquerel} for $^{99}$Mo`s daughter isotope, $^\text{99m}$Tc) within a few hours.






\section*{Funding}
This study was financed in part by the Conselho Nacional de Desenvolvimento Cientifico e Tecnológico (CNPq) (Grants No. 405143/2021-4 and 140941/2023-1), the Fundação de Amparo à Pesquisa do Estado do Rio Grande do Sul (FAPERGS) (Grant No. 21/2551-0002027-0), and the Coordenação de Aperfeiçoamento de Pessoal de Nível Superior – Brasil (CAPES) (Grant No. 88887.620985/2021-00). Javier Resta López is supported by the Generalitat Valenciana, Spain, under grant agreement CIDEGENT/2019/058.

\section*{CRediT authorship contribution statement}

\textbf{Bruno Silveira Nunes:} Writing – review \& editing, Writing – original draft, Validation, Software, Methodology, Investigation, Formal analysis, Conceptualization and Data curation. \textbf{Samara Prass dos Santos:} Writing – review \& editing. \textbf{Roger Pizzato Nunes:} Writing – review \& editing, Software. \textbf{Cristian Bon\c{t}oiu:} Writing – review \& editing. \textbf{Mirko Salomón Alva Sánchez:} Writing – review \& editing, Supervision, Funding acquisition, Project administration. \textbf{Ricardo Elgul Samad:} Writing – review \& editing,  Funding acquisition, Conceptualization. \textbf{Nilson Dias Vieira Jr.:} Writing – review \& editing, Funding acquisition, Conceptualization. \textbf{Guoxing Xia:} Writing – review \& editing. \textbf{Javier Resta López:} Writing – review \& editing. \textbf{Alexandre Bonatto:} Writing – review \& editing, Supervision, Conceptualization, Funding acquisition, Project administration.

\section*{Declaration of competing interest}
The authors declare that they have no known competing financial interests or personal relationships that could have appeared to influence the work reported in this paper.

\section*{Data Availability Statement}

The data that support the findings of this study are available from the corresponding authors upon reasonable request.

\section*{Acknowledgements}

The authors acknowledge computing resources provided by LNCC’s SDumont HPC system (project LPA-FARMA), and the N8 Centre of Excellence in Computationally Intensive Research (N8 CIR), provided and funded by the N8 research partnership and EPSRC (Grant No. EP/T022167/1).




\bibliographystyle{elsarticle-num}
\bibliography{references_short}






\end{document}